\documentclass[aps,prb,twocolumn,groupedaddress,showpacs]{revtex4-1}

\usepackage{graphicx}
\usepackage{hyperref}
\usepackage{color}

\begin{document}


\title{Interplay between energy dissipation and reservoir-induced thermalization in nonequilibrium quantum nanodevices}


\author{Fabrizio Dolcini${}^{1,2}$}
\email[]{fabrizio.dolcini@polito.it}
\author{Rita Claudia Iotti${}^1$}
\author{Fausto Rossi${}^1$}
\affiliation{${}^1$
Department of Applied Science and Technology, Politecnico di Torino, 
I-10129 Torino, Italy \\
${}^2$ CNR-SPIN, I-80126 Napoli, Italy}


\date{\today}

\begin{abstract}

A solid state electronic nanodevice is an intrinsically open quantum system, exchanging both energy with the host material and carriers with connected reservoirs. Its out-of-equilibrium behavior is determined by a non-trivial interplay between electronic dissipation and decoherence induced by inelastic processes within the device, and the coupling of the latter to metallic electrodes. We propose a unified description, based on the density matrix formalism, that accounts for both these aspects, enabling to predict various steady-state as well as ultrafast nonequilibrium phenomena, nowadays experimentally accessible. More specifically, we derive a generalized density-matrix equation, particularly suitable for the design and optimization of a wide class of electronic and optoelectronic quantum devices. The power and flexibility of this approach is demonstrated with the application to a photoexcited triple-barrier nanodevice.

\end{abstract}


\maketitle

\section{Introduction}\label{s-I}

The never-ending technological progress in solid state meso- and nanosystems, such as quantum dots\cite{QDs} and graphene-based devices,\cite{graphene} enables one to observe a large variety of phase-coherence phenomena,\cite{PC} where the wavelike properties of electrons become apparent. 
In any realistic quantum device, however, phase coherence  is hindered by  the  inelastic scattering that electrons experience within the host material (phonons, photons, plasmons, etc.) as well as by the coupling to external reservoirs;
such phenomena may cause energy dissipation, decoherence, and carrier thermalization, thereby affecting the nanodevice behavior. Relevant examples in current quantum-device physics and technology include charge as well as spin decoherence in state-of-the-art semiconductor macroatoms and carbon-based materials, with a special focus on quantum-information and spintronic applications.\cite{decoherence}

Within such general framework, two different and well established fields may be identified. On the one hand, the progressive reduction of the space- and time-scales of  electronic and optoelectronic devices forces one to replace the traditional Boltzmann picture\cite{Jacoboni89} with genuine quantum approaches based, e.g., on the density-matrix formalism,\cite{Rossi11} on the Green's function theory,\cite{Haug07} and on the Wigner-function picture.\cite{Frensley90}
On the other hand, a realistic description of new generation quantum-transport devices requires to extend the  Landauer-B\"uttiker treatment of the mesoscopic transport regime,\cite{Buettiker92} in order to include energy-dissipation and decoherence phenomena.

Any solid state electronic nanodevice is an intrinsically open quantum system, exchanging both energy with the host material and carriers with connected reservoirs. Its out-of-equilibrium behavior is determined by a complex interplay between electronic dissipation and decoherence induced by inelastic processes within the device, and the coupling of the latter to metallic electrodes. 
Aim of this paper is to provide a unified  framework, based on the density-matrix formalism, able to describe such highly non-trivial behavior, thus enabling one to predict various steady-state as well as ultrafast nonequilibrium phenomena, nowadays experimentally accessible. 

The article is organized as follows: after recalling the density-matrix formalism for quantum-device modeling (Sec.~\ref{s-FQDM}),  we  derive in Sec.~\ref{s-PTS}  a generalized density-matrix equation, particularly suitable for the design and optimization of a wide class of electronic and optoelectronic quantum devices.
In Sec.~\ref{s-PE} the power and flexibility of the proposed approach is demonstrated with the application to a photoexcited triple-barrier nanodevice.
Finally, in Sec.~\ref{s-SC} we discuss the results and draw a few conclusions.

\section{Density-matrix formalism applied to quantum device modeling}\label{s-FQDM}

The crucial interplay between electronic phase-coherence and dissipation versus decoherence phenomena in semiconductor bulk and nanostructures is often described through the electronic single-particle density matrix\cite{Rossi11}
\begin{equation}\label{rho}
\rho_{\alpha_1\alpha_2} = 
\left\langle
\hat c^\dagger_{\alpha_2} \hat c^{ }_{\alpha_1} \right\rangle\ ,
\end{equation}
where $\alpha$ spans the set of noninteracting carrier states (typically given by the scattering states related to the device potential profile) and 
$\hat c^\dagger_\alpha$ ($\hat c^{ }_\alpha$) denote the corresponding creation (annihilation) operators.\cite{note-rho} In Eq.(\ref{rho}),  $\langle \ldots \rangle$ includes the average over environment and reservoirs' degrees of freedom. 
As discussed in Ref.~[\onlinecite{Rossi02}], by adopting a number of well established approximation schemes ---including the well known Markov limit as well as the mean-field approximation--- the equation of motion for the electronic single-particle density matrix (\ref{rho}) can be written as\cite{Rossi11}  
\begin{equation}\label{SBE}
{d \rho_{\alpha_1\alpha_2} \over dt} = {\epsilon_{\alpha_1} - \epsilon_{\alpha_2} \over i \hbar} \rho_{\alpha_1\alpha_2} + \left. \frac{d \rho_{\alpha_1\alpha_2}}{dt} \right|_{\rm env} \, .  
\end{equation}
In Eq.(\ref{SBE}), the first term on the r.h.s. accounts for the coherent evolution, possibly including elastic scattering processes, dictated by the non-interacting single-particle Hamiltonian 
\begin{equation}\label{Hcirc}
\hat{H}^\circ = \sum_\alpha \epsilon_\alpha \hat{c}^\dagger_\alpha \hat{c}^{ }_\alpha
\end{equation} 
($\epsilon_\alpha$ denoting the energy levels corresponding to the single-particle states $\alpha$). In contrast, the second term on the r.h.s. of Eq.(\ref{SBE}) encodes dissipation and decoherence processes, arising from the energy exchange between the carriers and the host material; henceforth we shall refer to such term as the carrier-environment (env) coupling.
Equation~(\ref{SBE}) applies to a broad variety of problems, a remarkable example being the semiconductor Bloch equations.\cite{Rossi02}   \\

Importantly, the degree of accuracy of Eq.~(\ref{SBE}) is closely related to an appropriate choice of its last term. Indeed, oversimplified phenomenological treatments can lead, for instance, to a violation of the positive-definite character of the density-matrix operator 
\begin{equation}\label{hatrho}
\hat\rho = \sum_{\alpha_1\alpha_2} \vert\alpha_1\rangle \rho_{\alpha_1\alpha_2} \langle\alpha_2\vert\ ,
\end{equation} 
which is a mandatory prerequisite of any quantum-mechanical time evolution. 
To this end, a general prescription is to express various energy-relaxation coupling mechanisms via suitable Lindblad superoperators.\cite{Lindblad76} In this way, the last term in (\ref{SBE}) can be written in operatorial form as 
\begin{equation}\label{env-term}
{d \hat\rho \over d t}\biggl|_{\rm env} = 
\sum_s \left(
\hat A^{}_s \hat\rho \, \hat   A_s^\dagger
- 
{1 \over 2} \left\{\hat A_s^\dagger \hat A_s^{}, \hat\rho\right\} 
\right)\ ,
\end{equation} 
where $\hat{A}^{}_s$ denotes the  Lindblad superoperator related to the $s$-th interaction mechanism with the host material. \\

\begin{figure}
\centering
\includegraphics*[width=7cm]{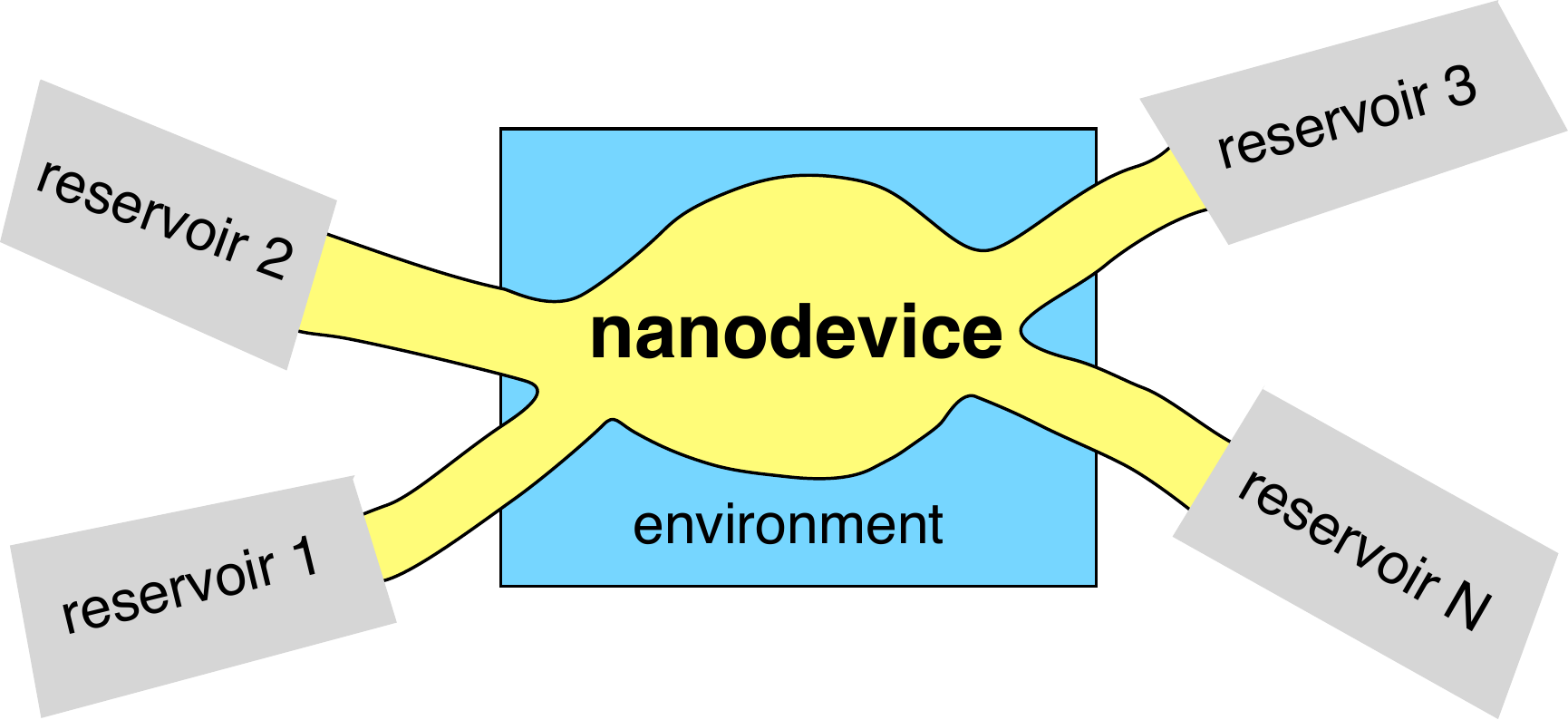}
\caption[]{(Color online)
A nanodevice as an intrinsically open quantum system, which can exchange energy with the host material (environment), and carriers with external reservoirs. Its nonequilibrium properties are determined by the non trivial interplay between these two types of interactions.}
\label{Fig1} 
\end{figure}

Besides the environment, an electronic or optoelectronic nanodevice is also coupled to   metallic electrodes, i.e., to macroscopic charge reservoirs. 
This makes any realistic nanodevice an intrinsically open quantum system, exchanging energy with the environment and carriers with various reservoirs, as schematically depicted in Fig.~\ref{Fig1}. 
However, these two interaction channels cannot be treated on equal footing. 
Indeed, the coupling to reservoirs cannot be accounted for via a term like Eq.~(\ref{env-term}), which necessarily implies that $d({\rm tr}\hat\rho)/dt=0$. Different approaches are thus needed to account for the fact that, while $\hat\rho$ remains a positively-defined operator, its trace  is in general {\it not} preserved. 

In the coherent regime, where inelastic coupling with the environment is negligible, the Landauer-B\"uttiker formalism\cite{Buettiker92} provides a successful description of the system steady state in the presence of external reservoirs. However, when inelastic scattering mechanisms are present, this treatment becomes inadequate.  

An alternative strategy  is  the Wigner-function approach,\cite{Frensley90} where dissipation and decoherence phenomena due to the environment are usually described within the relaxation-time approximation, whereas the presence of the reservoirs is encoded via suitable spatial boundary conditions. However,   such a classical-like approach  has been recently shown\cite{Taj06Rosati13} to lead to negative carrier probability densities. 

A general and physically consistent description, including dissipation and decoherence phenomena due to the open character of the electronic system, is therefore still lacking and will be proposed in the next section. 

\section{Proposed theoretical scheme}\label{s-PTS}

In this section we propose a unified and physically reliable framework for the description of open quantum systems, which enables one to account for dissipation and decoherence due to the energy exchange with the environment as well as for carrier transfer from/into the reservoirs. In Sec.~\ref{s-PE} we shall show that this approach is particularly suitable for the design and optimization of new-generation quantum devices with open spatial boundaries, corresponding, e.g., to the case of a semiconductor nanodevice inserted into an electric circuit, schematically depicted as the $N$ reservoirs in Fig.~\ref{Fig1}. 

The explicit form of the desired system-reservoir coupling superoperator should fulfill three basic requirements. It should (i)~preserve the positivity of the density-matrix operator~$\hat\rho$; (ii)~not induce phase coherence on the carrier system, due to the thermal (i.e., fully incoherent) nature of the external reservoirs; (iii)~reduce to the standard injection-loss structure in the semiclassical limit ($\rho_{\alpha_1\alpha_2} = f_{\alpha_1} \delta_{\alpha_1\alpha_2}$),\cite{Rossi11} i.e.
\begin{equation}\label{sm}
{d f_\alpha \over d t}\biggl|_{\rm res}
= S_\alpha - \Gamma^{\rm res}_\alpha f_\alpha = -\Gamma^{\rm res}_\alpha (f_\alpha-f^{\rm res}_\alpha) \ ,
\end{equation}
where  $\Gamma^{\rm res}_\alpha$ denotes the inverse timescale for $f_\alpha$ to reach the steady-state distribution $f_{\alpha}^{\rm res}$ in the absence of carrier-environment coupling.\\

Adopting a fully operatorial notation, the proposed extension of the density-matrix equation~(\ref{SBE}), compatible with all the above requirements, reads 
\begin{equation}\label{new}
{d \hat\rho \over dt} = \frac{1}{i\hbar} \left[H^\circ \, ,\hat{\rho} \right] \,  + \left. \frac{d \hat\rho}{d t} \right|_{\rm env} \,+ \left. \frac{d \hat\rho}{d t} \right|_{\rm res} \,
\end{equation}
where
\begin{equation}
\label{res-term}
{d \hat\rho \over d t}\biggl|_{\rm res}
=
\sum_{j=1}^{N} \sum_{k^j}
\left(
\hat{B}_{k^j}^{} \hat\rho^\circ \hat{B}_{k^j}^{\dagger}
- 
{1 \over 2} \left\{ \hat{B}_{k^j}^{} \, \hat{B}_{k^j}^{\dagger}, \hat\rho\right\}\right) \, .
\end{equation}
Here 
\begin{equation}\label{hatrhocirc}
\hat\rho^\circ = \sum_{j=1}^N \sum_{k^j} \vert k^j \rangle f^{\circ}_{k^j} \langle k^j \vert\ ,
\end{equation}
is the density-matrix operator encoding the distribution of the incoming reservoir electrons, with $f^{\circ}_{k^j}$ denoting the equilibrium or quasiequilibrium carrier distribution of the $j$-th reservoir,\cite{note-env} whereas
\begin{equation}\label{hatAk}
\hat B_{k^j} = \vert \alpha_{k^j} \rangle \sqrt{\Gamma^{\rm res}_{\alpha_{k^j}}} \langle k^j \vert  
\end{equation}
are Lindblad-like superoperators describing the coupling between the generic free-particle state $\vert k^j \rangle$ incoming from the $j$-th reservoir and the corresponding single-particle scattering states $\vert \alpha_{k^j} \rangle$ of the nanodevice. Differently from the genuine Lindblad form in~(\ref{env-term}), describing the energy exchange with the environment, the first term on the r.h.s. of Eq.~(\ref{res-term}) involves the density-matrix operator $\hat\rho^\circ$ of the external reservoirs. This feature, in sharp contrast with the superoperator in (\ref{env-term}), makes Eq.~(\ref{new}) inhomogeneous, implying that ${\rm tr}\hat\rho$ is not conserved, as expected in a system where the number of particles may change with time. Nevertheless, the positive-definite character of $\hat\rho$ is ensured by the Lindblad-like form of the proposed coupling term. It is also worth noticing that in our approach  the local (i.e., classical-like) boundary-condition treatment of the system-reservoir  interaction employed in the Wigner formulation\cite{Frensley90} has been replaced by a non-local quantum-mechanical  coupling. \\

Let us finally rewrite the operatorial equation (\ref{new}) within the single-particle basis $\alpha$. By introducing the density-matrix operator 
\begin{equation}\label{hatrhores}
\hat\rho^{\rm res} = \sum_{j} \sum_{k^j} |\alpha_{k^j}\rangle f^{\circ}_{\alpha_{k^j}} \langle \alpha_{k^j}| \, \neq \hat\rho^\circ\ ,
\end{equation} 
and by inserting Eqs.~(\ref{hatrhocirc}) and~(\ref{hatAk}) into Eqs.(\ref{new})-(\ref{res-term}) one gets
\begin{eqnarray}\label{new_entries}
{d \rho_{\alpha_1\alpha_2} \over dt} &=& {\epsilon_{\alpha_1} - \epsilon_{\alpha_2} \over i \hbar} \rho_{\alpha_1\alpha_2}  +\left. \frac{d \rho_{\alpha_1 \alpha_2}}{d t} \right|_{\rm env} -\nonumber \\
& &
- {\Gamma^{\rm res}_{\alpha_1} + \Gamma^{\rm res}_{\alpha_2} \over 2} \left(\rho_{\alpha_1\alpha_2}-\rho^{\rm res}_{\alpha_1\alpha_2}\right) \ ,
\end{eqnarray}
where the compact notation $\alpha_{k^j} \rightarrow \alpha$  has been employed. \\

In order to illustrate the implications of the system-reservoir coupling [second line in Eq.~(\ref{new_entries})], let us start by analyzing the mesoscopic regime, i.e. the case where  inelastic scattering mechanism are negligible and thus the environment coupling term in Eq.~(\ref{new_entries}) can be omitted. Then, the steady-state (ss) solution of Eq.~(\ref{new_entries}) is easily obtained as 
\begin{equation}\label{rhoss}
\rho^{\rm ss}_{\alpha_1\alpha_2} = \rho^{\rm res}_{\alpha_1\alpha_2} = \, f^\circ_{\alpha_1} \,\delta_{\alpha_1 \alpha_2}\ ,
\end{equation}
and the Landauer-B\"uttiker result\cite{Buettiker92} is recovered: the steady-state single-particle density matrix $\rho^{\rm ss}_{\alpha_1\alpha_2}$ of the device is diagonal, thus preventing any phase-coherence transfer between reservoirs and device. Its diagonal values coincide with the distribution $f^{\circ}_\alpha$ of the carriers injected from the reservoirs. Notice that such steady-state solution $\rho^{\rm ss}_{\alpha_1\alpha_2}$ is independent of the value of the device-reservoir coupling constants $\Gamma^{\rm res}_\alpha$. However, our approach goes beyond  the steady-state regime and enables one to address ultrafast-dynamics phenomena, nowadays accessible via time-resolved optical experiments. In particular, if one initially ``prepares'' a device electron in a coherent superposition of the single-particle states $|\alpha\rangle$, characterized by a device density matrix $\overline{\rho}_{\alpha_1\alpha_2}$ at $t = 0$, its subsequent time evolution is given by:
\begin{equation}\label{te}
\rho_{\alpha_1\alpha_2}(t) = \rho^{\rm ss}_{\alpha_1\alpha_2} +
\left(
\overline{\rho}_{\alpha_1\alpha_2}-\rho^{\rm ss}_{\alpha_1\alpha_2}\right)
e^{{(\Sigma^*_{\alpha_1}-\Sigma^{ }_{\alpha_2}) t \over i \hbar}}\ ,
\end{equation} 
where 
\begin{equation}\label{Sigma}
\Sigma_\alpha = \epsilon_\alpha + i {\hbar\Gamma^{\rm res}_\alpha / 2}
\end{equation}
can be regarded to as the device single-particle self-energy $\epsilon_\alpha$ ``dressed'' by an imaginary life-time contribution induced by the device-reservoir interaction. Thus Eq.~(\ref{te}) enables one to interpret the coupling $\Gamma_\alpha=\Gamma_{\alpha_{k^j}}$ as the inverse time-scale over which the diagonal density-matrix element $\rho_{\alpha \alpha}$ reaches the steady state value $\rho^{\rm ss}_{\alpha \alpha}$, if only injection from the $j$-th reservoir were present. Furthermore, the time evolution in (\ref{te})  shows the interplay between phase coherence  and reservoir-induced dissipation versus decoherence processes. Indeed in Eq.~(\ref{te})   the diagonal contributions ($\alpha_1 = \alpha_2$)  describe the population transfer and therefore energy dissipation, whereas the off-diagonal contributions ($\alpha_1 \ne \alpha_2$) exhibit a temporal decay of the inter-state polarizations, leading to decoherence. \\
It is also worth pointing out the difference between the treatment of the system-reservoir coupling presented here and the Bloch-Wangsness-Redfield (BWR) density-matrix formalism.\cite{BWR,Petruccione} BWR equations, which are typically applied to the analysis of nuclear spin relaxation, quantum optics and molecular dynamics, are not necessarily of Lindblad form (except for some specific cases e.g. in laser physics\cite{JCP2000}), so that the positivity of the density matrix is not guaranteed; moreover, the trace of the density matrix is preserved, in agreement with the fact that  those systems do not typically exchange particles with external reservoirs. In contrast, for the device-reservoir coupling discussed here the positivity of $\hat\rho$ is ensured by the Lindblad-like form of the superoperator (\ref{res-term}), whereas the ${\rm tr}\{\hat\rho\}$ is not preserved, in agreement with the fact that a nanodevice typically exchanges electrons with the contacted metallic  electrodes.\\

 Let us now include the presence of inelastic interaction mechanisms with the environment (phonons, photons, plasmons, etc.). We denote by  $\rho_{\alpha_1 \alpha_2}^{\rm env}$  the steady-state density matrix entries when a nanodevice is disconnected from
the external circuit (i.e. in the absence of external reservoirs). A first flavor of the effect of the environment can be obtained by   assuming that   
its coupling to the system can be described via a relaxation-time approximation. Then, the thermalization process to $\rho_{\alpha_1 \alpha_2}^{\rm env}$ can be expressed in terms of partially phenomenological coupling constants 
\begin{equation}
\Gamma_{\alpha}^{\rm env}=\sum_s \Gamma_{\alpha}^{{\rm env},s}\ ,
\end{equation}
where $\Gamma_{\alpha}^{{\rm env},s}$ is operatively defined as the inverse timescale that the system would take to reach the steady-state solution, if only the $s$-th interaction mechanism with the environment were present. \\
When the system is further connected to the reservoirs, the steady-state solution  within the relaxation-time picture turns out to be
\begin{equation}\label{sss2}
\rho^{\rm ss}_{\alpha_1\alpha_2} = 
{
(
\Gamma^{\rm env}_{\alpha_1}
+
\Gamma^{\rm env}_{\alpha_2}
)
\rho^{\rm env}_{\alpha_1\alpha_2}
+ 
(
\Gamma^{\rm res}_{\alpha_1}
+
\Gamma^{\rm res}_{\alpha_2}
)
\rho^{\rm res}_{\alpha_1\alpha_2}
\over
(
\Gamma^{\rm env}_{\alpha_1}
+
\Gamma^{\rm env}_{\alpha_2}
)
+
(
\Gamma^{\rm res}_{\alpha_1}
+
\Gamma^{\rm res}_{\alpha_2}
)
} \ .
\end{equation}
As we can see, the above  steady-state solution is the result of a non-trivial interplay between dissipation versus decoherence processes induced by the external reservoirs, and those induced by various inelastic processes within the device active region. 
It is then quite natural to identify two limiting regimes. For 
$\Gamma^{\rm env}_\alpha \gg \Gamma^{\rm res}_\alpha$
the effect of inelastic interaction processes is dominant over the reservoirs one, and     one obtains
$\rho^{\rm ss}_{\alpha_1\alpha_2} = \rho^{\rm env}_{\alpha_1\alpha_2}$, i.e., the effect of the scattering processes within the device is such to maintain the electron gas in thermal equilibrium with the host material (environment-controlled regime). In contrast, for
$\Gamma^{\rm env}_\alpha \ll \Gamma^{\rm res}_\alpha$
the effect of inelastic scattering is negligible, and the mesoscopic steady-state solution in (\ref{rhoss}) is recovered, i.e., the latter is essentially determined  by the external reservoirs (reservoir-controlled regime).\\

One may consider to reword these regimes in terms of ratio between two  timescales $\tau_0$ and $\tau$, where $\tau_0$ denotes the timescale the electron distribution  thermalizes to the environment when the reservoirs are disconnected (mean free time), and $\tau$ describes the timescale an electron takes to experience the distribution of all the electrodes when the environment is disconnected (device flight time). 
Then, the environment-controlled regime is obtained for $\tau_0 \ll \tau$, whereas the reservoir-controlled regime is obtained for $\tau \ll \tau_0$.
We notice that such reduction to  only two timescales is in general not straightforward,  despite $\Gamma^{{\rm res},s}_\alpha$ and $\Gamma^{\rm env}_\alpha$ are defined as inverse timescales. Indeed   the specific value of $\Gamma_{\alpha}^{{\rm env},s}$ heavily depends on the nanodevice geometry, as well as on the specific interaction mechanism considered. Secondly, in a multi-terminal device there are in principle various timescales related to the reservoirs. Finally, every state~$\alpha$ contributes with its own  timescale, and the states   weight differently to the density-matrix operator $\hat\rho$. Nevertheless, if one considers  a two-terminal device, and only one scattering mechanism for the environment, an appropriately weighted average  $\langle \ldots \rangle_\alpha$  may lead to identify  
$\tau_0 = 1/\langle \Gamma_\alpha^{\rm env}\rangle_\alpha$  and $\tau = 1/\langle\Gamma_\alpha^{\rm res}\rangle_\alpha$. \\

To conclude this section we point out that the simplified description of environment-induced coupling via the relaxation-time approximation mentioned above does not allow one to provide a quantitative (i.e., parameter-free) evaluation. To this end one has to describe the environment-induced time evolution via the Lindblad superoperator in (\ref{env-term}). Indeed, as shown recently,\cite{Taj09} by adopting an alternative Markov procedure, it is possible to perform a fully microscopic derivation of the Lindblad superoperators in (\ref{env-term}); the latter, written within the basis~$\alpha$, involve off-diagonal scattering rates expressed via a generalized Fermi's golden rule.
The highly non-trivial structure of such scattering superoperators does not allow for an analytical treatment of the problem; numerical solutions are needed and will be presented for a realistic nanodevice in the next section.  

\section{A prototypical example}\label{s-PE}

\begin{figure}
\centering
\includegraphics*[width=8cm]{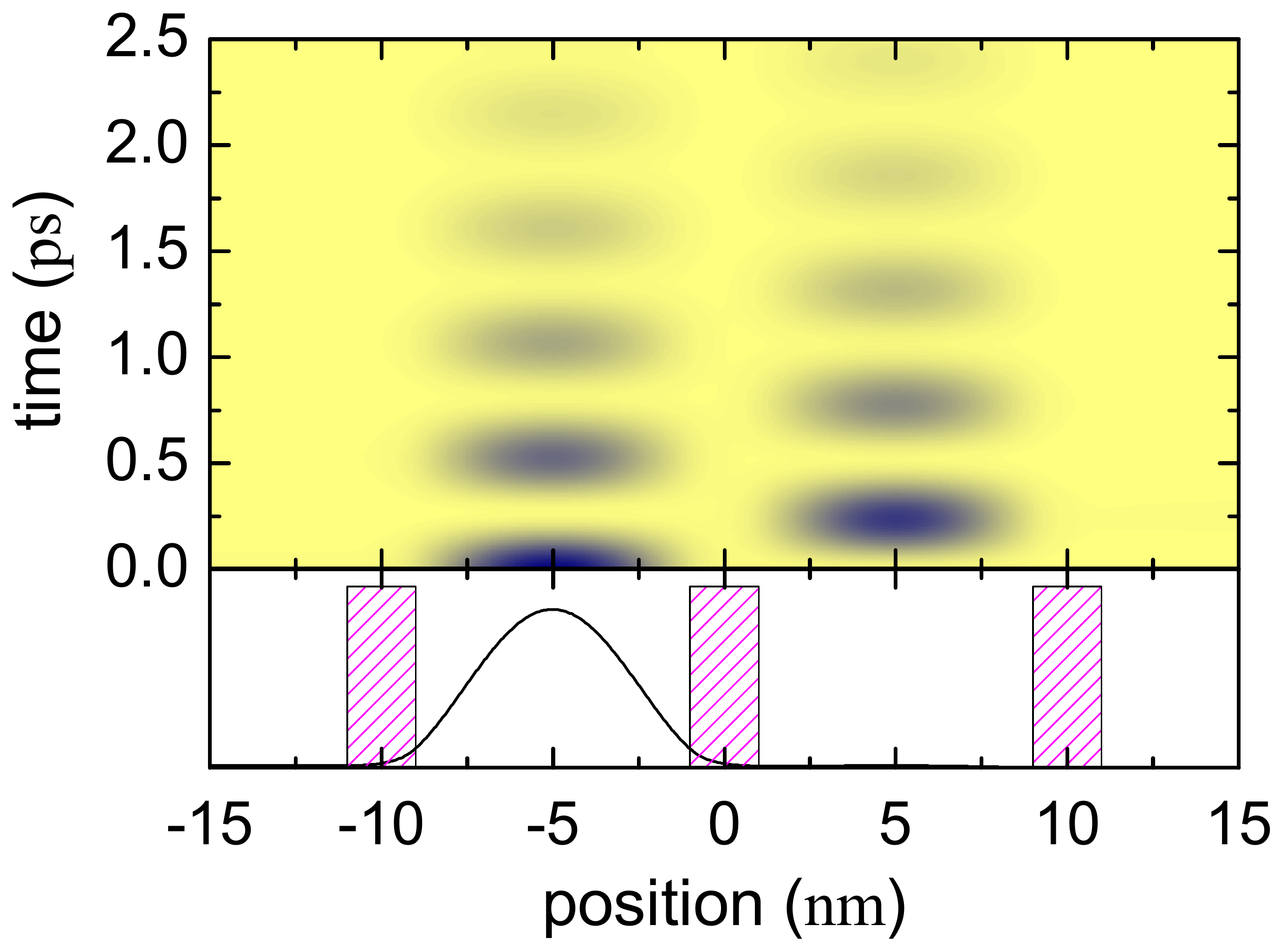}
\caption[]{(Color online)
Time evolution of the photoexcited carrier density across a realistic GaAs/AlGaAs triple-barrier nanodevice (well width $a_w = 8$\,nm, barrier width $a_b = 2$\,nm and height $V_0 = 300$\,meV).
At $t$ = 0 photoexcited carriers are fully localized in the left well (lower panel).
The subsequent transient dynamics (upper panel), corresponding to the solid-curve result in Fig.~\ref{Fig3}b, clearly shows an interwell tunneling dynamics characterized by a strong interplay between phase coherence and energy dissipation and decoherence processes
(see text).}
\label{Fig2}       
\end{figure}

\begin{figure}
\centering
\includegraphics*[width=8cm]{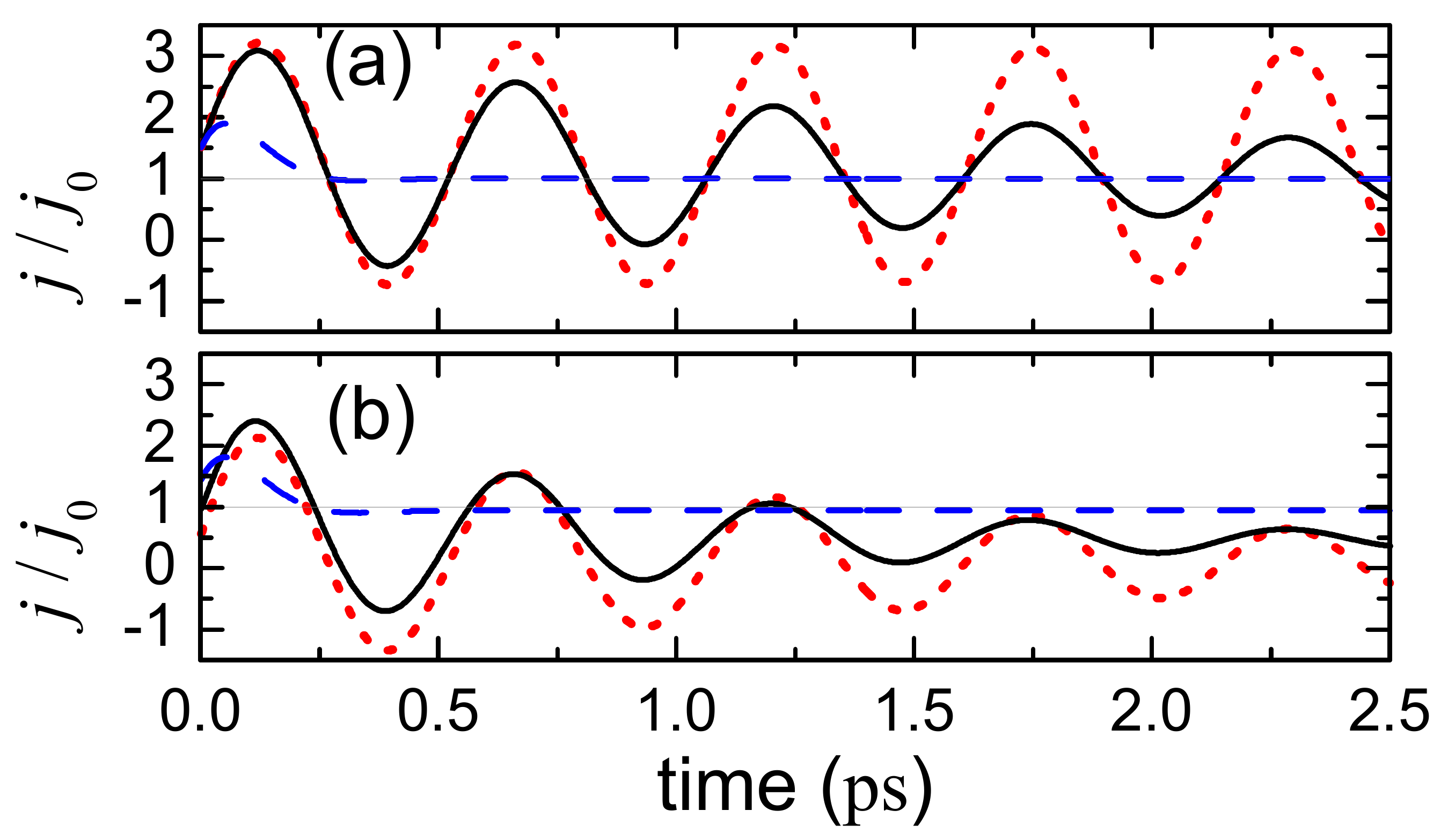}
\caption[]{(Color online)
Ultrafast electro-optical response of a realistic GaAs/AlGaAs triple-barrier nanodevice (see lower panel in Fig.~\ref{Fig2}) sandwiched between its electric contacts:
current density $j$ across the central barrier (in units of its Landauer-B\"uttiker value $j_0$) 
as a function of time corresponding to an initial coherent superposition of the two resonant electronic states of the triple-barrier structure ($\epsilon_1 \simeq 48$\,meV, $\epsilon_2 \simeq 56$\,meV) and to a quasiequilibrium carrier injection from the left contact only (total charge density $n \simeq 7 \times 10^{16}$\,cm$^{-3}$) in the mesoscopic limit (a), and in the presence of inelastic scattering processes (b), for three different values of the effective device length 
($l = 50$\,nm (dashed curves), $l = 1 \, \mu$m (solid curves), and $l = 20 \, \mu$m (dotted curves)).
}
\label{Fig3}       
\end{figure}

In order to show the power and flexibility of the proposed quantum-transport formalism, we have applied this simulation strategy to the investigation of dissipation-versus-decoherence phenomena in a realistic GaAs-based nanodevice. To this end, we have performed a numerical solution of the density-matrix equation (\ref{new_entries}) by employing a fully three-dimensional description of the electronic states $\alpha$ obtained via a standard transfer-matrix calculation within the conventional envelope-function and effective-mass approximations.\cite{Rossi11}
More specifically, we have considered a symmetric GaAs/AlGaAs triple-barrier structure (whose profile is sketched in the lower panel of Fig.~\ref{Fig2}), connected to two electrodes, tailored in such a way to allow for a significant interwell tunneling corresponding to an energy splitting between its resonant states of about $8$\,meV. This allows one to ``prepare'' the electronic system in a coherent superposition of these resonant states such to localize the electrons within the left well, and then to investigate the subsequent coherent charge and current oscillations between the two wells.
This scenario is fully confirmed by the simulated experiment reported in Fig.~\ref{Fig2}: at $t$ = 0 photoexcited carriers are fully localized in the left well (lower panel);
the subsequent transient dynamics (upper panel) clearly shows an interwell tunneling dynamics characterized by a strong interplay between phase coherence and energy relaxation and decoherence processes.

As discussed   in Ref.[\onlinecite{Kuhn94}], the above mentioned ``preparation'' of the electronic system can be experimentally realized by means of a properly tailored ultrafast interband optical excitation, by replacing our symmetric triple barrier structure with a biased asymmetric one. However, while the use of an asymmetric structure is experimentally crucial in order to photoexcite carriers in one well only, our simplified choice of a symmetric profile still retains the relevant physical features we are interested in. 

For the nanodevice under investigation, the primary source of energy dissipation and decoherence induced by the environment is carrier-optical phonon scattering in the host material. We have treated  microscopically such scattering, according to the general prescription given in Ref.~[\onlinecite{Taj09}] via the Lindblad scattering superoperator in (\ref{env-term}), whose matrix elements within the basis~$\alpha$ involve off-diagonal scattering rates expressed via a generalized Fermi's golden rule.
The mean free time $\tau_{0}$, resulting from such microscopic calculation of the carrier-phonon scattering, is of the order of $1$\,ps, corresponding to a scattering mean free path $l_0$ of the order of $1 \, \mu$m.

In order to investigate the non-trivial interplay between steady-state transport properties and ultrafast optical excitations, given the single-particle density matrix $\rho_{\alpha_1\alpha_2}$ obtained as a numerical solution of Eq.~(\ref{new_entries}), we have evaluated the average spatial carrier density
\begin{equation}\label{nz}
n(z) = \sum_{\alpha_1\alpha_2} \phi^*_{\alpha_1}\!(z) \,\phi^{ }_{\alpha_2}\!(z) 
\rho_{\alpha_2\alpha_1}
\end{equation}
as well as the corresponding average current density
\begin{equation}\label{Jz}
j(z) = -\frac{\hbar}{2 m^*} \sum_{\alpha_1\alpha_2}
\left[
i \, \phi^*_{\alpha_1}\!(z) \, \partial_z \phi_{\alpha_2}\!(z) + \textrm{h.c.} \right]
\rho_{\alpha_2\alpha_1}
\end{equation}
along the nanostructure growth direction $z$.
Here, $\phi_\alpha(z) = \langle z \vert \alpha \rangle$ is the wavefunction corresponding to the single-particle state $\alpha$, $m^*$ is the carrier effective mass, and h.c. denotes the hermitian conjugate.

As an initial condition of our simulated experiments we have chosen a single-particle density matrix $\overline{\rho}_{\alpha_1\alpha_2}$ given by the steady-state solution $\rho^{\rm ss}_{\alpha_1\alpha_2}$ plus a contribution $\Delta\rho_{\alpha_1\alpha_2}$ corresponding to the coherent superposition of the resonant states just mentioned, which mimics the effect of an ultrafast interband excitation.\cite{Kuhn94}
Starting from this initial condition, we have investigated the value of the current density $j$ in (\ref{Jz}) across the central barrier as a function of time for different values of the device effective lengths in the absence (mesoscopic limit) and presence of inelastic scattering processes, with varying the coupling with reservoirs. To this end, we have expressed such coupling in terms of an effective device length $l$, such that $\Gamma^{\rm res}_{\alpha}=|v_\alpha|/l$, where $v_\alpha$ is the carrier group velocity. 

As shown in Fig.~\ref{Fig3}, in the mesoscopic limit (a) the initial current oscillations ---unambiguous fingerprint of a coherent interwell tunneling dynamics--- are progressively suppressed due to the presence of the external reservoirs, and, as expected, the dissipation-induced decoherence time-scale decreases with the effective device length $l$. In the presence of inelastic scattering processes (b) the mesoscopic-limit scenario in (a) is partially modified:
as expected, in the short-device limit ($l \ll l_0$, dashed curve) inelastic scattering plays a minor role (compared to the action of the reservoirs), while in the opposite limit ($l \gg l_0$, dotted curve) the latter dominates, giving rise to a damping of the coherent oscillations as well as to a suppression of the steady-state current; in the intermediate regime $l \sim l_0$ (solid curves) one deals with a non-trivial interplay between reservoir- and environment-induced dissipation versus decoherence, resulting in a faster damping of the oscillations. The solid-curve result in Fig.~\ref{Fig3}b corresponds to the transient dynamics shown in the upper panel of Fig.~\ref{Fig2}.

\section{Discussion and Conclusions}\label{s-SC}

In the present article, we have proposed a conceptually unified and physically reliable framework for the   description of energy dissipation and decoherence in open quantum systems, able to overcome a few crucial drawbacks and limitations of current quantum-device modeling strategies.
More specifically, we have derived a generalized density-matrix equation, particularly suited for the design and optimization of new-generation semiconductor nanodevices with open spatial boundaries. The power and flexibility of this new modeling paradigm has been confirmed via fully three-dimensional simulated experiments. A few remarks are now in order. \\

In formulating the device-reservoir coupling term in~(\ref{res-term}), a crucial assumption is that the reservoirs are always in thermal or quasi-thermal equilibrium, so that no signature of phase coherence can arise from the reservoirs. This assumption is motivated by the fact that the typical system of interest for the proposed analysis is a semiconductor nanodevice, for which the reservoirs are metallic contacts characterized by a negligible coherence length. Thus,  in most of the experimentally relevant conditions, coherence effects induced by the reservoirs can fairly be neglected. It is worth pointing out that this assumption may not apply to other types of physical systems, such as atoms interacting with an electromagnetic vacuum,\cite{AEV} coherent phonons,\cite{CP} and microcavity polaritons,\cite{MP} where the environment and/or the reservoirs may induce coherence on the system (see also below). For these cases the application of the proposed coupling model may be questionable.
 
Under the discussed validity conditions, the proposed Eq.(\ref{new}) for $\hat{\rho}$ enables one to both recover the correct results in known limits and to describe the interplay between the effects of the reservoirs and the environment on the single-particle electronic density matrix. Within such scheme, our formulation has an autonomous   logical consistency, and does not rely on a microscopic derivation. 
However, an open question is whether it is possible to microscopically derive the Lindblad-like operators $B_{k^j}$ in Eq.(\ref{new}). Here we would like to discuss the issues involved in such a derivation. 
In quite general terms, the effective description of the carrier subsystem via the single-particle density matrix (\ref{rho}) may also be regarded to as the result of a reduction procedure, i.e., a suitable statistical average of the whole (device + reservoirs) system dynamics over non-relevant degrees of freedom, which include in this case the reservoirs as well. 
One could thus start from a global system Hamiltonian characterized by independent device and reservoir degrees of freedom, and describe the device-reservoir coupling via tunnel-like interaction terms. In general, such terms induce a highly non-trivial many-body dynamics, characterized by a number of phase coherence phenomena affecting both the device and the reservoir subsystem. 
Treating the device-reservoir coupling to lowest order in perturbation theory within the Markov approximation, an equation can thus be obtained for the global density matrix; if the Markov limit just mentioned is performed adopting a time-symmetrization scheme recently proposed,\cite{Taj09} the resulting equation for the global evolution is always of Lindblad form.
Nevertheless, given such global time evolution, several issues play an important role in deriving the equation for the reduced density matrix. This is realized by performing the average over non-relevant degrees of freedom previously mentioned, and therefore crucially depends on the problem under examination.
For the case of coupling of electrons to a phononic {\it environment}, it has been shown that, by taking the trace over the phononic degrees of freedom, the environment Lindblad operators can be microscopically derived.\cite{Taj09} However, for the case of {\it reservoirs}, an additional problem arises, due to the fact that the device exchanges particles with them, leading to a non-preservation of the trace of $\hat\rho$.
For these reasons, the microscopic derivation of the proposed reservoir Lindblad-like superoperator (\ref{res-term})  remains a challenging open question, which goes beyond the purpose of the present paper.\\

Finally, it is worth discussing how the ultrafast oscillations of the simulated photocurrent reported in Fig.~\ref{Fig3} can be measured. On the one hand, a  direct measurement of this inter-well photocurrent is hardly accessible experimentally, since such current oscillations across the central barrier have no counterpart in the external circuit.
On the other hand, it is possible to perform a time-resolved detection of the electromagnetic field produced by the charge oscillations of Fig.~\ref{Fig2}; indeed, the latter   generates an electromagnetic signal within the terahertz spectral region proportional to the time derivative of the photocurrent. Such a terahertz emission has first been observed from asymmetric double quantum-well structures,\cite{Roskos92} thus opening the way to the field of terahertz spectroscopy in semiconductors.\cite{Rossi02}




\begin{thebibliography}{99}
\bibitem{note-rho}
Its diagonal elements ($\alpha_1 = \alpha_2$) correspond to the
single-particle carrier distribution of the semiclassical
Boltzmann theory, while the off-diagonal contributions ($\alpha_1
\neq \alpha_2$), also referred to as interstate polarizations,\cite{Rossi02} describe quantum-mechanical phase coherence between the single-particle states $\alpha_1$ and $\alpha_2$.
\bibitem{note-env}
Indeed, for solid-state quantum devices the external circuit can be described as a number of free-carrier reservoirs (corresponding to one or more electric contacts), and its quasiequilibrium density-matrix operator is diagonal in the plane-wave basis.
\end{thebibliography}

\begin{acknowledgments}

F.D. acknowledges financial support from FIRB 2012 project ÓHybridNanoDevÓ (Grant No.RBFR1236VV).

\end{acknowledgments}

\end{document}